\documentclass[aps,pra,twocolumn,groupedaddress,showpacs,eqsecnum]{revtex4}

\usepackage{graphicx}

\begin{document}

%\preprint{}

\title{Measuring the elements of the optical density matrix}


\author{K. L. Pregnell}
\email{K.Pregnell@mailbox.gu.edu.au}
\author{D. T. Pegg}
\email{D.Pegg@sct.gu.edu.au}

%\thanks{}

\affiliation{School of Science, Griffith University, Brisbane 4111, Australia}


\date{\today}

\begin{abstract}
Most methods for experimentally reconstructing the quantum state of light involve determining a quasiprobability
distribution such as the Wigner function. In this paper we present a scheme for measuring individual density matrix
elements in the photon number state representation. Remarkably, the scheme is simple, involving two beam splitters and
a reference field in a coherent state.
\end{abstract}

\pacs{03.65.Wj, 42.50.-p}
%\keywords{}

\maketitle


\section{Introduction}

It is now well established that the quantum state of light can be measured. The first experimental evidence of this
\cite{Smithey} followed the work of Vogel and Risken \cite{Vogel}, where it was shown that the Wigner function could be
reconstructed from a complete ensemble of measured quadrature amplitude distributions. The authors of \cite{Smithey}
measured the quadrature distributions using balanced homodyne techniques. In the case of inefficient homodyne
detectors, a more general s-parameterized quasiprobability distribution is obtained resulting in a smoothed Wigner
function. In either case, to obtain the quasiprobability phase space distribution from the measured data a rather
complicated inverse transformation is required.

Novel techniques which avoid this transformation are aimed at measuring the quasiprobability distribution more
directly. This can be achieved, for example, in unbalanced homodyne counting experiments \cite{Banaszek,Wallentowitz},
where a weighted sum of photocount statistics are combined to obtain a single point in the phase space distribution.
The entire distribution is then obtained by scanning the magnitude and phase of the local oscillator over the region of
interest while repeating the photon counting at each point. Perhaps the most direct method of obtaining a
quasiprobability distribution is to use heterodyne \cite{Shapiro} or double homodyne \cite{Walker, Noh} detection
techniques where the $Q$ function is measured. The $Q$ function is related to the Wigner function through a convolution
with a gaussian distribution which effectively washes out many of the interesting quantum features. It is possible to
recover these features by deconvoluting the $Q$ function, however this requires multiplying by an exponentially
increasing function thereby introducing a crucial dependence on sampling noise \cite{Vogel2}. Other non-tomographic
state reconstruction schemes are proposed in \cite{9a} for fields in a cavity, in \cite{10a} for trapped atoms and in
\cite{11a} where use of a Schr\"{o}dinger-cat state probe is suggested. Further discussion of such techniques can be
found in the recent review by Welsch and Vogel \cite{12a}, the book by Leonhardt \cite{13a} and in \cite{14a}.

A different approach has been suggested by Steuernagel and Vaccaro \cite{SV}, who have proposed an interesting
nonrecursive scheme to measure not the quasiprobability distribution, but rather the density operator in the photon
number basis. The scheme is relatively direct in that only a finite number of different measurements are required to
determine each matrix element. However, the major disadvantage of this scheme is that determination of each matrix
element, $\rho_{MN}$, requires the preparation of a two-state superposition of $|N\rangle$ and $|M\rangle$ for use as a
probe field. Not only has the preparation of such fields not yet been achieved, but also the experiment requires a
change of the probe field for the measurement of each matrix element.

In this paper we investigate an alternative non-recursive scheme to that of Steuernagel and Vaccaro that also measures
the individual density matrix elements in the photon number basis. Remarkably we find that this can be achieved with a
single reference field that can be in an easily prepared coherent state. The only changes to the reference field needed
to measure all the matrix elements are simple phase shifts. The integral transformation required in the tomographic
technique is avoided and is replaced by a summation of only four easily measurable probabilities. We find that this
technique is particulary suited for measurements of low intensity states, such as that used in \cite{Lvovsky}, where it
offers some simplifications over the tomographic methods.

\section{Measurement Technique}
The density matrix element $\rho_{MN}$ in the number state representation of a density operator $\hat\rho$ is given by
\begin{equation}
\rho_{MN}=\textrm{Tr}(\hat\rho|N\rangle\langle{M}|).\label{1a}
\end{equation}
This can be compared with the probability for an outcome event $e$ of a measurement on a system in state $\hat\rho$,
which is given from general quantum measurement theory \cite{Hel} by
\begin{equation}
P(e)=\textrm{Tr}[\hat\rho\widehat\Pi(e)],\label{2a}
\end{equation} where $\widehat\Pi(e)$ is the element of a probability operator measure (POM) associated with the event $e$.
Comparing (\ref{1a}) and (\ref{2a}) suggests that if we could find a POM element equal to the operator
$|N\rangle\langle{M}|$ then we could find the matrix element $\rho_{MN}$ simply by measuring the probability $P(e)$.
This of course is not possible as the probability must be between zero and one but the matrix element need not even be
real. However if we could synthesize the operator $|N\rangle\langle{M}|$ by a linear combination of different POM
elements then we could find the matrix element from the same linear combination of the associated measurable
probabilities. We adopt this \emph{operator synthesis} approach in this paper.

\begin{figure}
\includegraphics[width=0.40\textwidth]{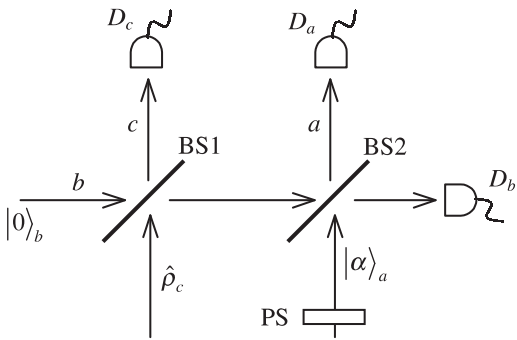}
\caption{Apparatus for measuring the density matrix elements of light. BS1 and BS2 are beam splitters. The field to be
measured and a reference field in a coherent state are in the input modes $c$ and $a$ of BS1 and BS2 respectively. A
vacuum is in the input mode $b$ and photon counters are in the output modes. Phase shifter PS adjusts the phase of the
coherent state.}\label{fig1}
\end{figure}

The proposed measurement technique uses the arrangement shown in Fig.~\ref{fig1}. It consists of two symmetric beam
splitters labelled BS1 with input modes $b$ and $c$ and BS2 with input modes $b$ and $a$. BS2 is a 50/50 beam splitter
but we keep the transmission to reflection coefficient ratio for BS1 general for now. In the output modes are
photodetectors $D_{a}$, $D_{b}$ and $D_{c}$. The input fields in modes $b$ and $a$ are respectively in a vacuum state
$|{0}\rangle_b$ and a coherent state $|\alpha\rangle_a$. The optimum value of the amplitude of the coherent state will
be discussed later. The field to be measured, in state $\hat\rho_{c}$, is in the input mode $c$ of BS1. In the entry
port of BS2 is a phase shifter PS capable of altering the phase of the coherent state, thereby changing the argument of
the coherent state amplitude. We let the amplitude of the coherent state be $\alpha =|\alpha|\exp(i\varphi)$ at the
entry of BS2, that is the argument $\varphi$ of $\alpha$ incorporates the phase shift. We let $\varphi$ be a function
of two numbers $\beta$ and $j$, that is $\varphi$ = $\varphi(\beta,j)$, that will be specified later.

For simplicity, we assume that the distance between the beam splitters is an integer number of wavelengths of the
light, which allows us to ignore the evolution of the light, which is just a phase shift, between the beam splitters.
If this is difficult experimentally, the discrepancy can be offset by an adjustment of the phase shifter. The complete
time evolution operator is then $\widehat R_2\widehat R_{1}$ where $\widehat R_2$ and $\widehat R_1$ are the unitary
operators for the action of beam splitters BS2 and BS1. Let $e$ = $(n_{a}, n_{b}, n_{c})$ be the event that
photodetectors $D_{a}$, $D_{b}$ and $D_{c}$ register $n_{a}, n_{b}$ and $n_{c}$ photocounts respectively. The
probability for this event is
\begin{eqnarray}
P_{\beta{j}}(e)=&\textrm{Tr}_{abc}(\widehat R_2\widehat
R_{1}\hat\sigma\,\widehat R_{1}^{\dagger}\widehat R_{2}^{\dagger}
|{n_a}\rangle_{a\,a}\langle{n_a}|\otimes|{n_b}\rangle_{b\,b}\langle{n_b}|
\nonumber
\\
&\mbox{}\otimes|{n_c}\rangle_{c\,c}\langle{n_c}|),\label{3a}
\end{eqnarray}
where we have written the combined density operator for the three input fields as
\begin{equation}
\hat\sigma=|{\alpha}\rangle_{a\,a}\langle{\alpha}|\otimes|{0}\rangle_{b\,b}
\langle{0}|\otimes\hat\rho_{c}\label{4a}
\end{equation}
and, as the subscripts imply, the trace is over the state spaces for all three modes. The subscript $\beta j$ on the
probability in (\ref{3a}) is to show explicitly that the probability is a function of the argument $\varphi(\beta,j)$
of $\alpha$, that is, it is a function of the setting of the phase shifter. Using the cyclic property of the trace we
can write (\ref{3a}) as
\begin{equation}
P_{{\beta}j}(e)=\textrm{Tr}_c\left[{\hat\rho_c\widehat\Pi_{\beta{j}}
(e)}\right],\label{5a}
\end{equation}
where
\begin{equation}
\widehat\Pi_{\beta{j}}(e)=|{q}\rangle_{c\,c}\langle{q}|\label{6a}
\end{equation}
with
\begin{equation}
|q\rangle_c={}_b\langle 0| \widehat{R}_1^\dagger\,{}_a\langle
\alpha|\widehat{R}_2^\dagger|{n_a}\rangle_a|{n_b}\rangle_b|{n_c}\rangle_c.
\label{7a}
\end{equation}
From (\ref{5a}) we see that $\widehat\Pi_{\beta{j}}(e)$ is an element of the POM for the measuring device that
comprises all of the arrangement depicted in Fig.~\ref{fig1} except for the field to be measured. In general the
elements of $\widehat\Pi_{\beta{j}}(e)$ are not necessarily orthogonal in that $\widehat\Pi_{\beta
j}(e)\widehat\Pi_{\beta j}(e')$ is not necessarily zero for $e\ne e'$. The origin of the non-orthogonality is the
introduction of the two reference modes $a$ and $b$. The effect of these two ancillary modes is to cube the
dimensionality of the system space. In considering the measuring apparatus to consist of everything in Fig.~1 except
the state to be measured, we effectively reduce the apparatus to a single mode measuring device with many more POM
elements than the dimensionality of the single mode. This means that the POM elements cannot all be orthogonal to one
another.

Our aim is to find a linear combination of POM elements equal to the operator $|N\rangle\langle{M}|$. It is convenient
to write this operator as $|N\rangle\langle{N+\lambda}|$ and consider separately the cases where $\lambda$ is even and
odd. We examine first the case where $\lambda$ is even. Consider the particular detection event $e_{1}$ = $(\lambda/2,
\lambda/2, N)$ in which the photodetectors $D_a$, $D_b$ and $D_c$ detect $\lambda/2$, $\lambda/2$ and $N$ photocounts
respectively. As shown in the Appendix, this turns out to be the optimum detection event for $\lambda$ even. The
unitary operator $\widehat R_{1}$ for BS1 is given by \cite{BR}
\begin{equation}
\widehat{R}_1=\exp[i\eta(\hat c^\dagger \hat b+\hat b^\dagger \hat c)],\label{8a}
\end{equation}
where $\cos\eta = t$ and $\sin\eta = r$ are the transmission and
reflection coefficients of BS1. $\widehat R_{2}$ for the 50/50
beam splitter BS2 is a similar function of $\hat a$ and $\hat b$
with $\eta~=~\pi/4$. Using these expressions with $n_{a}$, $n_{b}$
and $n_{c}$ equal to $\lambda/2$, $\lambda/2$ and $N$, we find
from the Appendix that (\ref{7a}) becomes
\begin{equation}
|q\rangle_c=\sum_{n=N}^{N+\lambda} f_{n}\exp[-i(N+\lambda-n)\varphi(\beta,j)]|n\rangle_c\label{9a}
\end{equation}
where
\begin{equation}
f_{n}=\frac{(2i)^{-\lambda/2}t^{N}|\alpha|^{\lambda+N-n}(-ir)^{n-N}
\sqrt{n!}}{\exp(|\alpha|^2/2)[(\lambda+N-n)/2]![(n-N)/2]!\sqrt{N!}}
\label{10a}
\end{equation}
if $n - N$ is even and $f_{n}$ = $0$ if $n - N$ is odd. The POM element (\ref{6a}) for the detection event $e_{1}$
then becomes
\begin{equation}
\widehat\Pi_{\beta{j}}(e_{1})=\sum_{n,m=N}^{N+\lambda}f_n
f_m^*\exp{[i(n-m)\varphi(\beta,j)]}|n\rangle_{c\,c} \langle m|.
\label{11a}
\end{equation}
The terms with $n-m$ odd are all zero.

For $\lambda$ = $0$ we find from (\ref{11a}) that the POM element in (\ref{6a}) is just proportional to
$|{N}\rangle_{c\,c}\langle{N}|$, allowing us to find the diagonal elements $\rho_{NN}$ of the density matrix from the
probability of detecting the event $(0, 0, N)$.

For $\lambda$ even and non-zero, we let $\varphi(\beta,j)$ take particular values
\begin{equation}
\varphi({\beta},j)=\frac{\beta\pi}{\lambda}+\frac{2\pi j}{\lambda}\label{12a}
\end{equation}
and consider a modified measurement procedure in which $\beta$ is held constant but the value of $j$ is cycled so that
it takes all the integer values from $0$ to $({\lambda}/2)-1$ with equal probability. This measurement procedure will
have its own probability operator measure comprised of elements $\widehat\Pi_\beta(e)$. This will be different from our
previous POM with elements $\widehat\Pi_{\beta j}(e)$ because it describes a different measurement process. The POM
element for detecting the event $e_{1}$ by means of this cycling procedure will be given by
\begin{eqnarray}
\lefteqn{\widehat \Pi_{\beta}(e_{1})={\frac{2}{\lambda}}
\sum_{j=0}^{\lambda/2-1}\widehat\Pi_{{\beta}j}(e_{1})}\nonumber
\\
&&={\frac{2}{\lambda}}\sum^{N+\lambda}_{n,m=N}\{f_nf_m^* \exp{[i(n-m)({{\beta}{\pi}/\lambda})]}\nonumber
\\
&&\mbox{}\times\sum^{\lambda/2-1}_{j=0}\exp{[i(n-m)2{\pi}j/\lambda]}|n\rangle_{c\,c} \langle m|\}.\label{14a}
\end{eqnarray}
The associated probability can be obtained in practice from the occurrence frequency of the event $e_{1}$ as we cycle
through the values of $j$ with the experiment being repeated an equal number of times for each value of $j$. Because we
need only consider terms in (\ref{14a}) for which $n-m$ is even, we can take the factor involving the summation over
$j$ as zero unless $n-m$ is zero or $\pm\lambda$, in which case it equals $\lambda/2$. This gives us
\begin{eqnarray}
\widehat {\Pi}_{\beta}(e_{1})&=&\sum^{N+\lambda}_{n=N}|f_n|^2|n\rangle_{c\,c} \langle n|+[f_Nf_{N+\lambda}^*\nonumber
\\
&&\mbox{}\times\exp(-i\pi\beta)
|N\rangle_{c\,c} \langle N+\lambda|+h.c.].\label{15a}
\end{eqnarray}

By choosing different values for $\beta$, we obtain different cycling experiments, each with its own POM.
Experimentally this means cycling through a different set of phase settings. It is not difficult to see from
(\ref{15a}) that a linear combination of POM elements $\widehat\Pi_{\beta}(e_{1})$ with $\beta$ taking the values 0, 1,
1/2 and 3/2 is required to synthesize the operator $|N\rangle\langle N+\lambda|$. Specifically,
\begin{eqnarray}
\lefteqn{|N\rangle\langle N+\lambda|=}\nonumber\\
&&\frac{\widehat
{\Pi}_{0}(e_{1})-\widehat {\Pi}_{1}(e_{1})+i\left[\widehat
{\Pi}_{1/2}(e_{1})
-\widehat{\Pi}_{3/2}(e_{1})\right]}{4f_Nf_{N+\lambda}^*},\label{16a}
\end{eqnarray}
where
\begin{equation}
\label{16b}f_Nf^*_{N+\lambda}=t^{2N}(ir/2)^\lambda |a_0a_\lambda^*|{\lambda\choose\lambda/2}{N+\lambda\choose N}^{1/2}
\end{equation}
is a normalisation constant with $a_n=\langle n|\alpha\rangle$.

Taking the trace of the product of the density operator $\hat\rho_{c}$ of the field to be measured with both sides of
(\ref{16a}) gives the desired matrix element in terms of the measurable probabilities $P_{\beta}(e_{1})$ for detecting
the event $(e_{1})$:
\begin{eqnarray}
\lefteqn{\langle N+\lambda|\hat\rho_{c}|N\rangle=}\nonumber\\
&&\frac{P_{0}(e_{1})-P_{1}(e_{1})+i\left[
P_{1/2}(e_{1})-P_{3/2}(e_{1})\right]}{4f_Nf_{N+\lambda}^*}\label{17a}
\end{eqnarray}
for $\lambda$ non-zero and even. The complex conjugate of (\ref{17a}) is $\langle N|\hat\rho_{c}|N+\lambda\rangle$.

To find the density matrix element for $\lambda$ odd, we consider
the detection event $e_{2}$ = $[(\lambda +1)/2, (\lambda -1)/2,
N]$, which is shown in the Appendix to be the optimum detection
event for this case. A derivation similar to that above eventually
yields the associated POM element of the form
\begin{equation}
\widehat\Pi_{\beta{j}}(e_{2})=\sum_{n,m=N}^{N+\lambda}g_n g_m^*\exp{[i(n-m)\varphi(\beta,j)]}|n\rangle_{c\,c}\langle m|.
\label{18a}
\end{equation}
In (\ref{18a}) $n-m$ takes all integer values from $+\lambda$ to $-\lambda$. We consider a measurement procedure where
$\beta$ is held constant but $j$ takes all values from $0$ to $\lambda-1$ with equal probability. The POM element
$\widehat\Pi_{\beta}(e_{2})$ for detecting the event $e_{2}$ with this procedure will be given by
\begin{eqnarray}
\lefteqn{\widehat
\Pi_{\beta}(e_{2})={\frac{1}{\lambda}}
\sum_{j=0}^{\lambda-1}\widehat\Pi_{{\beta}j}(e_{2})}\nonumber
\\
&&={\frac{1}{\lambda}}\sum^{N+\lambda}_{n,m=N}\{g_ng_m^* \exp{[i(n-m)({{\beta}{\pi}/\lambda})]}\nonumber
\\
&&\times\sum^{\lambda-1}_{j=0}
\exp{[i(n-m)2{\pi}j/\lambda]}|n\rangle_{c\,c} \langle m|\}.\label{20a}
\end{eqnarray}
The factor involving the summation over $j$ is zero unless $n-m$ = $0$ or $\pm\lambda$ and is then equal to $\lambda$.
We find that the formula for the density matrix element in terms of the measurable probabilities $P_{\beta}(e_{2})$ for
detecting the event $e_{2}$ is, for $\lambda$ odd,
\begin{eqnarray}
\lefteqn{\langle N+\lambda|\hat\rho_{c}|N\rangle=}\nonumber\\
&&\frac{ P_0(e_{2})-P_1(e_{2})
+i\left[ P_{1/2}(e_{2})-P_{3/2}(e_{2})\right]}{4g_N
g_{N+\lambda}^*},\label{21a}
\end{eqnarray}
where
\begin{equation}
\label{22a} g_Ng_{N+\lambda}^*=it^{2N}(ir/2)^\lambda
|a_0a_\lambda^*|{\lambda\choose(\lambda-1)/2}{N+\lambda\choose
N}^{1/2}
\end{equation}
We note the number of phase settings required for each experiment for the odd-$\lambda$ case is about twice that
required for the even-$\lambda$ case with a similar value of $\lambda$. The number of experiments needed for the
odd-$\lambda$ case can, however, be reduced by a factor of two as follows. While measuring the probability for the
event $e_{2}$, we can also measure the probability for another event $e_{3}$ = $[(\lambda -1)/2, (\lambda +1)/2, N]$.
It is possible to show that the probability $P_{\beta}(e_{3})$ is equal to  $P_{\beta +1}(e_{2})$. Thus we only need
two experiments with different values of $\beta$ to obtain all four terms in the numerator of (\ref{21a}).

\section{Some Practical Considerations}

To illustrate the viability of our proposal we shall, in this section, show the extent to which physical imperfections,
such as a noisy local oscillator, detector inefficiency and dark counts, can be ignored or compensated for in a
practical experiment.

A more general measurement scheme would have an arbitrarily mixed reference state at the input of mode $a$ in Fig. 1.
That is $|\alpha\rangle_a$, with $\alpha=|\alpha|\exp(i\varphi)$ at the entry of BS2, would be replaced by a density
operator $\exp(i\widehat N_a\varphi)\hat\rho_a\exp(-i\widehat N_a\varphi)$. This could represent, as a specific
example, a noisy local oscillator. Following the derivation outlined above, it is straightforward to show the effect of
this is to replace the term $|a_0 a_\lambda^*|$ in equations (\ref{16b}) and (\ref{22a}) with the density matrix
element $\langle 0|\hat\rho_a|\lambda\rangle$. This is interesting because it shows how the density matrix element
$\langle N+\lambda|\hat \rho_c|N\rangle$ of the unknown field can be obtained directly from the density matrix element
$\langle 0|\hat \rho_a|\lambda\rangle$ of the reference field. Thus, provided we know the noise characteristics of the
reference state, we do not require it to be a noiseless coherent state, or indeed any particular pure state, to use it
to find the density matrix elements of the unknown field. So we find in general that a noisy local oscillator can
easily be used in the measurement scheme. A problem arises, however, if $\langle 0|\hat\rho_a|\lambda\rangle$ is
vanishingly small in that the measured probabilities will coincide with rare events as indicated by (\ref{17a}) and
(\ref{21a}). This is the case when phase diffusion in the local oscillator is prominent, effectively diagonalizing the
density matrix $\hat\rho_a$ and removing all phase information. This can be avoided if both the reference field and the
measured field, $\hat\rho_c$, are derived from a common source, a technique commonly exploited in experiments of this
kind.

Another practical issue concerns the extent to which inefficient photodetectors degrade the reliability of the measured
data. The effect of inefficient photodetectors is to make the measurement process uncertain. For a given detector
efficiency $\eta$, the probability of detecting $n$ photons, $p_n(\eta)$, is related to the probability of detecting
$m$ photons with a perfect detector, $p_{m}(1)$, by \cite{13}
\begin{equation}
\label{23}p_n(\eta)=\sum_{m=0}^\infty{n+m\choose n}\eta^n(1-\eta)^m p_{n+m}(1).
\end{equation}

To illustrate what effect inefficient photodetectors have on the outcome of the experiment, some numerical calculations
were performed for a low intensity coherent input state with a mean photon number of 0.5. An example of the results are
summarized in Tables \ref{tab1} and \ref{tab2},
\begin{table}
\caption{Truncated density matrix for a coherent state with a mean photon number of 0.5.} \label{tab1}
\begin{ruledtabular}
\begin{tabular}{ccccc}
    0.6065  &  0.4289  &  0.2145  &  0.0870  &  0.0336\\
    0.4289  &  0.3033  &  0.1517  &  0.0615  &  0.0238\\
    0.2145  &  0.1517  &  0.0759  &  0.0308  &  0.0119\\
    0.0870  &  0.0615  &  0.0308  &  0.0125  &  0.0048\\
    0.0336  &  0.0238  &  0.0119  &  0.0048  &  0.0019
\end{tabular}
\end{ruledtabular}
\end{table}
\begin{table}
\caption{Simulation of measured density matrix for a coherent
state with a mean photon number of 0.5 and detector inefficiency
$\eta=0.9$} \label{tab2}
\begin{ruledtabular}
\begin{tabular}{ccccc}
    0.6592  &  0.4195  &  0.1888  &  0.0692  &  0.0220\\
    0.4195  &  0.2967  &  0.1335  &  0.0489  &  0.0161\\
    0.1888  &  0.1335  &  0.0668  &  0.0244  &  0.0081\\
    0.0692  &  0.0489  &  0.0244  &  0.0100  &  0.0033\\
    0.0220  &  0.0161  &  0.0081  &  0.0033  &  0.0013
\end{tabular}
\end{ruledtabular}
\end{table} where the measured density matrix is displayed for a detector efficiency of $\eta=0.9$. The reference state
used in each simulation was a coherent state with a mean photon number of $|\alpha|^2=0.5$. As expected, as the
efficiency decreases the relative error in the individual matrix elements increases. Fortunately it is possible to
invert equation (\ref{23}) through a Bernoulli transformation and recover the exact probabilities from the detection
statistics with sufficiently good detectors \cite{13}. This would allow accurate reconstruction of the density matrix.

In addition, for weak fields in the quantum regime, with sufficiently long gating times, dead
times need not be significant. If dead times are significant, more sophisticated detection methods
are required for photon number discrimination, such as replacing each detector with a multiport
device such as described in \cite{14}.

So far we have not specified the value of $|\alpha|$ or $t/r$. The optimum values of these should maximize the
denominators of (\ref{17a}) and (\ref{21a}), thereby avoiding quotients of small numbers. We find that the optimum
value of $|\alpha|^{2}$ is $\lambda/2$ and that of $(t/r)^{2}$ is $2N/\lambda$. As these are optimum values only, they
need not be changed for the measurement of each matrix element and a reasonable compromise value should suffice, for
example, for weak fields where the spread of values of $N$ and $\lambda$ is not large.

While the method proposed in this paper can be used to measure any individual density matrix element, it is not
necessary to perform the same number of cycling experiments as matrix elements to find the density matrix. The matrix
elements $\langle N+\lambda|\hat\rho_{c}|N\rangle$ and their complex conjugates for all values of $N$ can be found from
the same four cycling experiments. Also many phase settings can be used as parts of different cycling experiments,
allowing further efficiencies. For example the setting $\varphi({\beta},j)$ = $\pi/2$ can be used for $\beta = 0$, $j =
1$, $\lambda = 4$ and $\beta = 0$, $j = 2$, $\lambda = 8$ as well as for $\beta = 1$, $j = 0$, $\lambda = 2$ and so on.

\section{Conclusion}
In this paper, we have extended the method of projection synthesis \cite{Projsyn}, in which a projector is synthesized
by use of an exotic reference state, to a more general technique of operator synthesis in which an operator is
synthesized by a linear combination of POM elements. This provides a nonrecursive method for measuring individual
density matrix elements of a light field. Remarkably, the technique is reasonably simple, involving only two beam
splitters and a reference field which can be in an easily-prepared coherent state. In particular, for states that can
be represented in a finite dimensional Hilbert space, this technique appears simpler than the tomographic methods in
that only a finite number of different measurements are required to ascertain the complete density matrix. We have
shown how detector inefficiency can be allowed for and have considered the effect of noise in the local oscillator. We
found that the local oscillator noise can be readily accounted for provided we know the corresponding mixed state
description of the local oscillator. Interestingly, our method allows the density matrix elements of the unknown field
to be obtained quite simply from the density matrix elements of a noisy local oscillator field, even when the unknown
field is in a pure state.

% If in two-column mode, this environment will change to single-column
% format so that long equations can be displayed. Use
%\begin{widetext}
%\end{widetext}

% figures should be put into the text as floats.
% Use the graphics or graphicx packages (distributed with LaTeX2e)
% and the \includegraphics macro defined in those packages.
% See the LaTeX Graphics Companion by Michel Goosens, Sebastian Rahtz,
% entire page. There is no need to do explicit centering.

% \begin{figure}
% \end{figure}

% Surround figure environment with turnpage environment for landscape
% figure
% \begin{figure}
% \end{figure}

\begin{acknowledgments}
DTP thanks the Australian Research Council for support.
\end{acknowledgments}

\appendix*
\section{}
In this appendix we derive the general form of $|q\rangle_c$
defined by (\ref{7a}):
\begin{equation}
\label{A1}|q\rangle_c={}_b\langle 0| \widehat{R}_1^\dagger\,{}_a\langle
\alpha|\widehat{R}_2^\dagger|{n_a}\rangle_a|{n_b}\rangle_b|{n_c}\rangle_c.
\end{equation}
With the unitary operator of a beam-splitter given by \cite{BR}
\begin{equation}
\label{A2}\widehat{R}=\exp[i\eta(\hat a^\dagger \hat b+\hat b^\dagger \hat a)],
\end{equation}
where $a$ and $b$ are the annihilation operators for the input field modes, it can be shown that the beam-splitter
transforms the corresponding creation operators, $a^\dag$ and $b^\dag$, and the double mode vacuum according to
\begin{eqnarray}
\label{A3}&\widehat{R}^{\dagger}\,\hat a^\dagger\widehat{R}=t\,\hat a^\dagger-ir\hat b^\dagger
\\
\label{A4}&\widehat{R}^{\dagger}\,\hat b^\dagger\widehat{R} =t\,\hat b^\dagger-ir\hat a^\dagger
\\
\label{A5}&\widehat{R}^{\dagger}|0\rangle_b|0\rangle_c=|0\rangle_b|0\rangle_c,
\end{eqnarray}
where $t$ and $r$ are the transmission and reflection coefficients of the beam-splitter. In the case of BS2, a 50/50
beam-splitter, $t=r=1/\sqrt{2}$. By writing $| n_a\rangle_a$ as $(\hat a^{\dag n_a}/\sqrt{n_a!})| 0\rangle_a$, and
similarly for $|n_b\rangle_b$, we obtain
\begin{equation}
\label{A7} \widehat{R}^{\dagger}_2|n_a\rangle_a|n_b\rangle_b=\frac{(\hat a^\dag-i\hat b^\dag )^{n_a}(\hat b^\dag-i\hat
a^\dag )^{n_b}}{2^{(n_a+n_b)/2}\sqrt{n_a!\,n_b!}}|0\rangle_a| 0\rangle_b
\end{equation}
and thus
\begin{eqnarray}
\label{A8}
\lefteqn{{}_a\langle\alpha|\widehat{R}^{\dagger}_2|n_a\rangle_a|n_b\rangle_b|
N\rangle_c\nonumber}\\ & &=\frac{(\alpha^*-i\hat b^\dag
)^{n_a}(\hat
b^\dag-i\alpha^*)^{n_b}}{2^{(n_a+n_b)/2}\exp(|\alpha|^2/2)\sqrt{n_a!\,n_b!}}|
0\rangle_b| N\rangle_c
\end{eqnarray}

Writing $| N\rangle_c$ as $(\hat c^{\dag N}/\sqrt{N!})|
0\rangle_c$ and using an equivalent form of (\ref{A3}-\ref{A5}),
we obtain
\begin{widetext}
\begin{eqnarray}
\label{A9}
\lefteqn{\widehat{R}^{\dagger}_1\,{}_a\langle\alpha|\widehat{R}^{\dagger}_2|n_a\rangle_a|n_b\rangle_b|
N\rangle_c\nonumber} \\& &=\frac{[\alpha^*-i(t\hat b^\dag-ir\hat
c^\dag)]^{n_a}[t\hat b^\dag-ir\hat c^\dag-i\alpha^*]^{n_b}(t\hat
c^\dag-ir\hat
b^\dag)^N}{2^{(n_a+n_b)/2}\exp(|\alpha|^2/2)\sqrt{n_a!\,n_b!N!}}|
0\rangle_b| 0\rangle_c
\end{eqnarray}
\end{widetext}
where we have left the transmission and reflection coefficients of BS1 as $t$ and $r$. Finally, projecting onto the
vacuum state in mode $b$ gives us
\begin{eqnarray}
|q\rangle_c&=&\frac{(-i)^{n_b}t^N [\alpha^*-r\hat
c^\dag]^{n_a}[\alpha^*+ r\hat
c^\dag]^{n_b}}{2^{(n_a+n_b)/2}\exp(|\alpha|^2/2)\sqrt{n_a!\,n_b!}}
|N\rangle_c\nonumber\\&=&\sum_{m=N}^{N+\lambda}
q_m(n_a,n_b)|m\rangle_c\label{A10b}
\end{eqnarray}
where $\lambda=n_a+n_b$. The specific notation
\begin{eqnarray}
\label{A10c}
f_m=q_m(\lambda/2,\lambda/2)\exp[i(N+\lambda-m)\varphi]\nonumber\\
g_m=q_m[(\lambda+1)/2,(\lambda-1)/2]\exp[i(N+\lambda-m)\varphi]
\end{eqnarray}
is used in the text. The explicit form of $q_m(n_a,n_b)$ is not actually needed. What is important is an expression for
$q_N(n_a,n_b)q_{N+\lambda}^*(n_a,n_b)$. This can be derived from (\ref{A10b}) by evaluating $q_N(n_a,n_b)$ and
$q_{N+\lambda}(n_a,n_b)$ separately to give
\begin{eqnarray}
\label{A10d}\lefteqn{q_N(n_a,n_b)q_{N+\lambda}^*(n_a,n_b)\nonumber}\\&=(-1)^{n_a}a_0a_\lambda^*
t^{2N}(r/2)^\lambda{\lambda\choose n_a}{N+\lambda\choose N}^{1/2}.
\end{eqnarray}
where $a_n=\langle n|\alpha\rangle.$ For a mixed reference state with density operator $\exp(i\widehat
N_a\varphi)\hat\rho_a\exp(-i\widehat N_a\varphi)$ at the entry of BS2, $a_0a_\lambda^*$ in (\ref{A10d}) is replaced by
$\langle 0|\hat\rho_a|\lambda\rangle\exp(-i\lambda\varphi)$.

It is not difficult to see that the modulus of (\ref{A10d}) is maximized when $n_a=\lambda/2$ if $\lambda$ is even and
when $n_a=(\lambda\pm1)/2$ if $\lambda$ is odd. Thus the quotients in (\ref{17a}) and (\ref{21a}) will have optimum
numerators and denominators for the values of $n_a$ and $n_b$ we have used in this paper.


\begin{thebibliography}{99}
\bibitem{Smithey} D. T. Smithey, M. Beck, M. G. Raymer, and
A. Faridani, Phys. Rev. Lett {\bf 70}, 1244 (1993); D. T. Smithey, M. Beck, J. Cooper, M. G. Raymer and A.
Faridani, Phys. Scr. {\bf T48}, 35 (1993).
\bibitem{Vogel} K. Vogel and H. Risken, Phys. Rev. A {\bf 40}, R2847 (1989).
\bibitem{Banaszek} K. Banaszek and K. W\'{o}dkiewicz, Phys. Rev. Lett. {\bf 76} 4344 (1996).
\bibitem{Wallentowitz} S. Wallentowitz and W. Vogel, Phys. Rev. A {\bf 53}, 4528 (1996)
\bibitem{Shapiro} J. H. Shapiro and S. S. Wagner, IEEE J. Quantum Electron. {\bf 20}, 803 (1984)
\bibitem{Walker} N. G. Walker and J. E. Carroll, Electron. Lett. {\bf}, 981 (1984).
\bibitem{Noh} J. W. Noh, A. Foug\`{e}res, and L. Mandel, Phys. Rev. Lett. {\bf 67}, 1426 (1991); Phys. Rev. A {\bf
45}, 424 (1992).
\bibitem{Vogel2} W. Vogel and J. Grabow, Phys. Rev. A {\bf 47}, 4227 (1993).
\bibitem{9a} P. J. Bardroff, E. Mayr, and W. P. Schleich, Phys. Rev. A {\bf 51}, 4963 (1995);
 P. J. Bardroff \textit{et al.}, Phys. Rev. A {\bf 53}, 2736 (1996);
 L. G. Lutterbach and L. Davidovich, Phys. Rev. Lett. {\bf 78}, 2547 (1997);
 M. S. Kim, G. Antesberger, C. T. Bodendorf, and H. Walther, Phys. Rev. A {\bf 58}, R65 (1998).
\bibitem{10a} D. Leibfried \textit{et al.}, Phys. Rev. Lett. {\bf 77}, 4281 (1996).
\bibitem{11a} D. G. Fischer and M. Freyberger, Opt. Commun. {\bf 159}, 158 (1999).
\bibitem{12a} D.-G. Welsch, W. Vogel, and T. Opatrn\'{y}, in \textit{Progress in Optics}, edited by E. Wolf
 (North-Holland, Amsterdam, 1999), Vol. 39, p. 63, and references therein.
\bibitem{13a} U. Leonhardt, \textit{Measuring the Quantum State of Light} (Cambridge University Press, Cambridge,
1997).
\bibitem{14a} M. S. Kim, G. Antesberger, C. T. Bodendorf, and H. Walther, Phys. Rev. A {\bf58}, R65 (1998);
 D. T. Pegg, L. S. Phillips, and S. M. Barnett, J. Mod. Opt. {\bf46}, 981 (1999); D.
T. Pegg and S. M. Barnett, J. Mod. Opt. {\bf46}, 1657 (1999); B. Rohwedder, L. Davidovich, and N. Zagury, Phys. Rev. A
{\bf 60}, 480 (1999); M. Mohmoudi, H. Tajalli, and M. S. Zubairy, J. Opt. B: Quantum Semiclass. Opt. {\bf 2}, 315
(2000); A. Zucchetti, S. Wallentowitz, W. Vogel, and N. P. Bigelow, Phys. Rev. A {\bf 61}, 011405 (2000); G. Nogues
\textit{et al.}, Phys. Rev. A {\bf 62}, 054101 (2000); M. F. Santos \textit{et al.}, Phys. Rev. A {\bf 63}, 033813
(2001); C. J. Villas-B\^{o}as, G. A. Prataviera, and M. H. Y. Moussa, Phys. Rev. A {\bf 64}, 065801 (2001); S. B.
Zheng, Commun. Theor. Phys. {\bf 36}, 213 (2001).
\bibitem{SV} O. Steuernagel and J. A. Vaccaro, Phys. Rev. Lett. {\bf 75},
3201 (1995).
\bibitem{Lvovsky} A. I. Lvovsky, H. Hansen, T. Aichele, O. Benson, J. Mlynek, and S. Schiller,
 Phys. Rev. Lett. {\bf 87}, 050402
(2001)
\bibitem{Hel} C. W. Helstrom, {\em Quantum Detection and Estimation Theory}
(Academic Press, New York, 1976).
\bibitem{BR} S. M. Barnett and P. M. Radmore, {\em Methods in Theoretical
Quantum
Optics} (Clarendon, Oxford, 1997).
\bibitem{13} C. T. Lee, Phys. Rev. A {\bf 48}, 2285 (1993); S. M. Barnett,
L. S. Phillips and D. T. Pegg, Opt. Commun. {\bf 158}, 45 (1998).
\bibitem{14} H. Paul, P. T\"{o}rm\"{a}, T. Kiss, and I. Jex, Phys. Rev. Lett. {\bf
76}, 2464 (1996).
\bibitem{Projsyn} S. M. Barnett and D. T. Pegg, Phys. Rev. Lett. {\bf 76},
4148 (1996).
\end{thebibliography}

\end{document}